\RequirePackage[]{graphicx} % Loaded here to show images in draft mode, too

\def\mode{1} % 1 is Arxiv

\if 1\mode
	\documentclass[journal]{article}
\fi

\if 2\mode
	\documentclass[num-refs]{wiley-article_final}	
\fi

\usepackage{etoolbox}
\newtoggle{MRM}
\newtoggle{ARXIV}

\newtoggle{SUPMATERIAL}
\togglefalse{SUPMATERIAL}

\if 1\mode
	\toggletrue{ARXIV}
	\togglefalse{MRM}
\fi

\if 2\mode
	\toggletrue{MRM}
	\togglefalse{ARXIV}
\fi

\iftoggle{MRM}{
	\usepackage[nomarkers, nolists]{endfloat}
}

\usepackage[utf8]{inputenc}
\usepackage[table]{xcolor}
\usepackage{adjustbox}
\usepackage{amsmath}
\usepackage{bm}
\usepackage{bbold}
\usepackage{authblk}
\usepackage{amssymb}
\usepackage{graphicx}
\usepackage{ifdraft}
\usepackage{marginnote}
\usepackage{pdfpages}
\usepackage{multirow}
\usepackage{tabularx,booktabs}
\usepackage{ifpdf}
\ifpdf
	\usepackage[]{hyperref}
\else
	\usepackage[dvipdfm]{hyperref}
\fi
\usepackage{tikz}
\usepackage[capitalise]{cleveref}

\usepackage{blindtext}

\usepackage[
per-mode=fraction,
separate-uncertainty,
retain-unity-mantissa=false,
product-units=power,
table-alignment=center,
detect-all,
binary-units=true,
exponent-product=\cdot,
exponent-base=10
]{siunitx}

\usepackage{color}

\usepackage{changes}
\newcommand{\stkout}[1]{\ifmmode\text{\sout{\ensuremath{#1}}}\else\sout{#1}\fi}
\setdeletedmarkup{\stkout{#1}}

\usepackage{mathtools}
\usepackage{gensymb}

\iftoggle{ARXIV}{\newcommand{\qed}{\hfill\blacksquare}}

\newtheorem{Lemma}{Lemma}
\newtheorem{Theorem}{Theorem}

\usepackage{placeins}

\title{{Rational Approximation of Golden Angles:\\Accelerated Reconstructions for Radial MRI}}

\newcommand{\authorA}{Nick Scholand}
\newcommand{\authorB}{Philip Schaten}
\newcommand{\authorC}{Christina Graf}
\newcommand{\authorD}{Daniel Mackner}
\newcommand{\authorE}{H. Christian M. Holme}
\newcommand{\authorF}{Moritz Blumenthal}
\newcommand{\authorG}{Andrew Mao}
\newcommand{\authorH}{Jakob Assländer}
\newcommand{\authorI}{Martin Uecker}

\newcommand{\affilA}{Institute of Biomedical Imaging, Graz University of Technology,
			Graz, Austria.}
\newcommand{\affilB}{Center for Biomedical Imaging, Department of Radiology,
	NYU School of Medicine, New York, USA.}
\newcommand{\affilC}{German Centre for Cardiovascular Research (DZHK),
	Partner Site Lower Saxony, Göttingen, Germany.}
\newcommand{\affilD}{Department of Pediatrics,
	The University of British Columbia, Vancouver, British Columbia, Canada.}
\newcommand{\affilE}{Department of Physics and Astronomy,
	The University of British Columbia, Vancouver, British Columbia, Canada.}
\newcommand{\affilF}{Institute for Diagnostic and Interventional Radiology,
	University Medical Center Göttingen, Göttingen, Germany.}
\newcommand{\affilG}{Center for Advanced Imaging Innovation and Research (CAI2R), Department of Radiology, NYU School of Medicine, New York, USA.}
\newcommand{\affilH}{Vilcek Institute of Graduate Biomedical Sciences, New York University School of Medicine, New York, USA.}
\newcommand{\affilI}{BioTechMed-Graz, Graz, Austria.}
\newcommand{\affilJ}{Cluster of Excellence “Multiscale Bioimaging: from Molecular Machines to Networks of Excitable Cells” (MBExC), University of Göttingen, Göttingen, Germany.}

\newcommand{\corMail}{scholand@tugraz.at}
\newcommand{\runHead}{Scholand et al.}

\iftoggle{ARXIV}{
		\author[1,2,3]{\authorA \thanks{\corMail}}
	}

\iftoggle{MRM}{
		\author[1,2,3]{\authorA}
	}

\author[1]{\authorB}
\author[1,4,5]{\authorC}
\author[1]{\authorD}
\author[1]{\authorE}
\author[1,6]{\authorF}
\author[2,7,8]{\authorG}
\author[2,7]{\authorH}
\author[1,3,6,9,10]{\authorI}

\affil[1]{\affilA}
\affil[2]{\affilB}
\affil[3]{\affilC}
\affil[4]{\affilD}
\affil[5]{\affilE}
\affil[6]{\affilF}
\affil[7]{\affilG}
\affil[8]{\affilH}
\affil[9]{\affilI}
\affil[10]{\affilJ}

\iftoggle{MRM}{
	\papertype{Note}
	\corraddress{\corAdress}
	\corremail{\corMail\\
			
			Submitted to MRM as note
			\bf{{Word count}}: Abstract: ???, Body: ???.}
	
	\runningauthor{\runHead}
}

\begin{document}

\maketitle

\iftoggle{ARXIV}{
	\vspace*{-0.7cm}
	\begin{center}
		\texttt{Submitted to Magnetic Resonance in Medicine}\\\vspace*{0.5cm}
		\textbf{Word count}: Abstract: 230, Main Body: $\sim$4750.
	\end{center}
}

\newpage

\begin{abstract}

\textbf{Purpose}:
To develop a generic radial sampling scheme that combines
the advantages of golden ratio sampling with simplicity
of equidistant angular patterns.
The irrational angle between consecutive spokes in golden ratio based
sampling schemes enables a flexible retrospective choice of 
temporal resolution, while preserving good coverage of k-space
for each individual bin.
Nevertheless, irrational increments prohibit precomputation
of the point-spread function (PSF), can lead to numerical problems, 
and require more complex processing steps.
To avoid these problems, a new sampling scheme based
on a rational approximation of golden angles (RAGA)
is developed.

\textbf{Methods:}
The theoretical properties of RAGA sampling are mathematically derived.
Sidelobe-to-peak ratios (SPR) are numerically computed and
compared to the corresponding golden ratio sampling schemes.
The sampling scheme is implemented in the BART toolbox
and in a radial gradient-echo sequence. Feasibility is
shown for quantitative imaging in a phantom
and a cardiac scan of a healthy volunteer.

\textbf{Results}:
RAGA sampling can accurately approximate golden ratio sampling
and has almost identical PSF and SPR. In contrast to
golden ratio sampling, each frame can be reconstructed
with the same equidistant trajectory using different
sampling masks, and the angle of each acquired spoke can
be encoded as a small index, which simplifies processing
of the acquired data.

\textbf{Conclusion}:
RAGA sampling provides the advantages of golden ratio
sampling while simplifying data processing, rendering
it a valuable tool for dynamic and quantitative MRI.

\vspace*{0.5cm}
\end{abstract}

\newpage

\section{Introduction}

Radial trajectories were the first sampling scheme used in MRI \cite{Lauterbur_Nature_1973}
and are now widely used in dynamic imaging
\cite{Glover_Magn.Reson.Med._1992,Larson_Magn.Reson.Med._2004,Benkert_Magn.Reson.Med_2018},
compressed sensing \cite{Block_Magn.Reson.Med._2007,Feng_Magn.Reson.Med._2014,Otazo_MagnResonMed_2015},
and quantitative MRI \cite{Block_IEEETrans.Med.Imaging_2009,Tran-Gia_PLOS_ONE_2015,Wang_Magn.Reson.Med._2018,Christodoulou_NatureBiomedicalEngineering_2018,Scholand_Magn.Reson.Med._2023}.
Radial trajectories have several advantages: They are ideally suited
for continuous acquisitions \cite{Rasche_Magn.Reson.Med._1995},
are robust to motion \cite{Glover_Magn.Reson.Med._1992,Katoh_J.Magn.Reson.Imaging_2006},
and they repeatedly acquire the k-space center which can be used for correction of
gradient imperfections \cite{Moussavi_Magn.Reson.Med._2013,Rosenzweig_Magn.Reson.Med._2018a},
self-navigation \cite{Rosenzweig_IEEETrans.Med.Imag._2020}, and calibration of coil sensitivities \cite{Uecker_Magn.Reson.Med._2014}. %FIXME: does ESPIRiT still count even though it is based on Cartesian?
For these reasons, radial acquisitions are increasingly used
in clinical applications including imaging of the heart, breast, abdomen, and brain
\cite{Block_J.KoreanSoc.Magn.Reson.Med._2014}.
Nevertheless, the reconstruction of data acquired with non-Cartesian
trajectories has a high computational cost, especially for iterative
reconstruction algorithms.
To reduce computation time, reconstruction techniques often use preprocessing steps
including prior interpolation/shifting of the data onto a Cartesian grid
\cite{Griswold_Magn.Reson.Med._2005,Seiberlich_Magn.Reson.Med._2007,Adluru_J.Magn.Reson.Imaging_2009}
or replacing the joint operation of interpolation and gridding by a convolution
with the point-spread function (PSF) in Toeplitz-based methods
\cite{Wajer__2001,Uecker_Magn.Reson.Med._2010,Baron_Magn.Reson.Med._2017}.

Radial sampling schemes differ in the angle between
spokes and in the overall temporal ordering of the acquisition of the spokes.
One of the simplest schemes is an equidistant angular pattern in which
the spokes are homogeneously distributed over a whole circle or half circle.
One advantage is that the acquired data is self-explanatory, i.e.
the angle between spokes can be derived from the total number of radial
projections. Another advantage is that only a finite number of
angular positions are sampled, which then allows
sharing of precomputed information about
quantities such as the PSF which may save memory or 
computation time.\\
While equidistant sampling is a good choice for static
imaging, other sampling schemes are preferable for dynamic imaging.
To improve spatio-temporal k-space coverage in
the presence of motion or dynamic contrast changes turn-based
acquisition schemes are used that reorder the spokes so that
consecutive spokes are separated by larger angles. 

A disadvantage of such schemes is that the temporal resolution
is determined by the temporal footprint of a single
turn. A more flexible solution is offered by golden ratio based
sampling \cite{Winkelmann_IEEETrans.Med.Imag._2007,Wundrak_IEEETransMedImag_2015}.
By choosing an irrational angle between spokes based on the golden ratio,
good k-space coverage is ensured simultaneously for different bin sizes. 
Temporal resolution can then be selected retrospectively.
Therefore, golden ratio based schemes are among the most common
radial sampling schemes \cite{Feng_J.Magn.Reson.Imaging_2022}.
Nevertheless, compared to equidistant angular sampling,
irrational angles
also have a disadvantage:
Every spoke samples a unique projection and 
this prevents 
sharing of precomputed information
%precomputation of sampling related quantities
such as the PSF % as used in Toeplitz-based gridding techniques
between different temporal frames of an acquired time series.

This work develops a new radial sampling scheme that combines the
advantages of equidistant angular patterns with the advantages
of golden ratio sampling.
The method, termed Rational Approximation of Golden Angles (RAGA),
exploits a generalized Fibonacci formulation
to compute a rational approximation of golden ratio based
angles. The concept of RAGA is introduced, and its theoretical
properties are described. Numerically and experimentally,
it is confirmed that RAGA sampling schemes closely match
the corresponding irrational sampling schemes. Feasibility
is demonstrated with phantom and in vivo scans.

\section{Theory}

\subsection{Rational Approximation of Golden Angles}

Golden ratio sampling as published by Winkelmann et al. \cite{Winkelmann_IEEETrans.Med.Imag._2007}
is based on the number known as the golden ratio, i.e.
\begin{align}
	\tau := \frac{\sqrt{5}+1}{2}~,
\end{align}
where the angle between consecutively acquired spokes is then given by 
\begin{align}
	\psi^1 := \frac{\pi}{\tau} \approx 111.245^{\circ}~.
\end{align}
Wundrak et al. \cite{Wundrak_IEEETransMedImag_2015,Wundrak_Magn.Reson.Med._2016_2}
generalized this concept also to smaller angles by defining the $N$-th golden ratio angles as
\begin{align}
	\psi^N := \frac{\pi}{\tau+N-1}~~~\text{and}~~~N = 1,2,\dots~.
	\label{Eq::psi_golden_ratio}
\end{align}
For $N > 1$, the $\psi^N$ are often called tiny golden angles.
All these angles are defined with respect to $\pi$, i.e. for a half-circle,
and are irrational numbers multiplied by $\pi$.
Note that $\psi^1$ is not the usual Golden Angle $\Phi\approx 137.508^{\circ}$
which is defined as
\begin{align}
	\Phi := 2\pi \left(1 -\frac{1}{\tau} \right) = 2\pi - 2 \psi^1~.
\end{align}

The basic idea behind RAGA is to find a suitable rational
approximation for the golden ratio angles $\psi^N$.
Besides Equation \eqref{Eq::psi_golden_ratio}, the golden ratio angle $\psi^N$
with index $N$ can also be derived from the generalized Fibonacci series \cite{Horadam_TheAmericanMathematicalMonthly_1961}
\begin{align}
	G_i^N := G_{i-1}^N + G_{i-2}^{N} ~~~~\text{with}~~~~ G_1^N := 1 ~~\text{and}~~ G_2^N := N~.
	\label{Eq::gen_fib}
\end{align}
The golden ratio angle $\psi^N$ is given by \cite{Wundrak_IEEETransMedImag_2015}
\begin{align}
	\psi^N= \pi\cdot \underset{i\rightarrow\infty}{\text{lim}}\frac{G_{i-1}^1}{G_i^N}~.
	\label{Eq::tiny-golden-angle-fib}
\end{align}
A rational approximation relative to $\pi$ can be obtained by
approximating $\psi^N$ with a finite order $i$ as
\begin{align}
	\psi^N \approx \psi_i^N := \pi\cdot \frac{G_{i-1}^1}{G_i^N}~.
	\label{Eq::raga_approx}
\end{align}
The higher the order $i$ the more accurate the approximation of $\psi^N$ becomes.
An overview about different angles for various approximation orders
$i$ is shown in Table~\ref{Tab:angles}.

\begin{table}[!ht]
	\renewcommand{\arraystretch}{1.2}
	\caption{$~$
		Rational approximation of various 
		golden ratio based angles $\psi^N$ for different approximation orders $i$.
		Each approximation $\psi_i^N$ is shown with their corresponding fraction 
		using generalized Fibonacci numbers.
		The bold numbers mark sampling schemes that fulfill
		the Nyquist criterion for a base resolution of 200,
		i.e. contain more than 314 projection angles.
		\label{Tab:angles} }
	\centerline{
	\begin{tabular}{ccc|c|c|c|c|c|c|c}
		& & \multicolumn{8}{c}{\textbf{Approximated Golden Ratio Angles}}\\
	 	& & & $\psi^{N=1}$ & $\psi^2$ & $\psi^3$ & $\psi^4$ & $\psi^5$ & $\psi^6$ & $\psi^7$ \\\cline{2-10}
		\multirow{24}{*}{\rotatebox[origin=c]{90}{\textbf{Approximation Order} $i$}}
% r! ./table1
& \multirow{2}{*}{2} & \cellcolor{gray!20} $\psi_{2}^N$ & \cellcolor{gray!20}     180.000$^\circ$ & \cellcolor{gray!20}      90.000$^\circ$ & \cellcolor{gray!20}      60.000$^\circ$ & \cellcolor{gray!20}      45.000$^\circ$ & \cellcolor{gray!20}      36.000$^\circ$ & \cellcolor{gray!20}      30.000$^\circ$ & \cellcolor{gray!20}      25.714$^\circ$ \\
 & & {\color{gray!70}$G_{1}^1 / G_{2}^N $} &  {\color{gray!70}   1 /    1} &  {\color{gray!70}   1 /    2} &  {\color{gray!70}   1 /    3} &  {\color{gray!70}   1 /    4} &  {\color{gray!70}   1 /    5} &  {\color{gray!70}   1 /    6} &  {\color{gray!70}   1 /    7} \\\cline{2-10}
 & \multirow{2}{*}{3} & \cellcolor{gray!20} $\psi_{3}^N$ & \cellcolor{gray!20}      90.000$^\circ$ & \cellcolor{gray!20}      60.000$^\circ$ & \cellcolor{gray!20}      45.000$^\circ$ & \cellcolor{gray!20}      36.000$^\circ$ & \cellcolor{gray!20}      30.000$^\circ$ & \cellcolor{gray!20}      25.714$^\circ$ & \cellcolor{gray!20}      22.500$^\circ$ \\
 & & {\color{gray!70}$G_{2}^1 / G_{3}^N $} &  {\color{gray!70}   1 /    2} &  {\color{gray!70}   1 /    3} &  {\color{gray!70}   1 /    4} &  {\color{gray!70}   1 /    5} &  {\color{gray!70}   1 /    6} &  {\color{gray!70}   1 /    7} &  {\color{gray!70}   1 /    8} \\\cline{2-10}
 & \multirow{2}{*}{4} & \cellcolor{gray!20} $\psi_{4}^N$ & \cellcolor{gray!20}     120.000$^\circ$ & \cellcolor{gray!20}      72.000$^\circ$ & \cellcolor{gray!20}      51.429$^\circ$ & \cellcolor{gray!20}      40.000$^\circ$ & \cellcolor{gray!20}      32.727$^\circ$ & \cellcolor{gray!20}      27.692$^\circ$ & \cellcolor{gray!20}      24.000$^\circ$ \\
 & & {\color{gray!70}$G_{3}^1 / G_{4}^N $} &  {\color{gray!70}   2 /    3} &  {\color{gray!70}   2 /    5} &  {\color{gray!70}   2 /    7} &  {\color{gray!70}   2 /    9} &  {\color{gray!70}   2 /   11} &  {\color{gray!70}   2 /   13} &  {\color{gray!70}   2 /   15} \\\cline{2-10}
 & \multirow{2}{*}{5} & \cellcolor{gray!20} $\psi_{5}^N$ & \cellcolor{gray!20}     108.000$^\circ$ & \cellcolor{gray!20}      67.500$^\circ$ & \cellcolor{gray!20}      49.091$^\circ$ & \cellcolor{gray!20}      38.571$^\circ$ & \cellcolor{gray!20}      31.765$^\circ$ & \cellcolor{gray!20}      27.000$^\circ$ & \cellcolor{gray!20}      23.478$^\circ$ \\
 & & {\color{gray!70}$G_{4}^1 / G_{5}^N $} &  {\color{gray!70}   3 /    5} &  {\color{gray!70}   3 /    8} &  {\color{gray!70}   3 /   11} &  {\color{gray!70}   3 /   14} &  {\color{gray!70}   3 /   17} &  {\color{gray!70}   3 /   20} &  {\color{gray!70}   3 /   23} \\\cline{2-10}
 & \multirow{2}{*}{6} & \cellcolor{gray!20} $\psi_{6}^N$ & \cellcolor{gray!20}     112.500$^\circ$ & \cellcolor{gray!20}      69.231$^\circ$ & \cellcolor{gray!20}      50.000$^\circ$ & \cellcolor{gray!20}      39.130$^\circ$ & \cellcolor{gray!20}      32.143$^\circ$ & \cellcolor{gray!20}      27.273$^\circ$ & \cellcolor{gray!20}      23.684$^\circ$ \\
 & & {\color{gray!70}$G_{5}^1 / G_{6}^N $} &  {\color{gray!70}   5 /    8} &  {\color{gray!70}   5 /   13} &  {\color{gray!70}   5 /   18} &  {\color{gray!70}   5 /   23} &  {\color{gray!70}   5 /   28} &  {\color{gray!70}   5 /   33} &  {\color{gray!70}   5 /   38} \\\cline{2-10}
 & \multirow{2}{*}{7} & \cellcolor{gray!20} $\psi_{7}^N$ & \cellcolor{gray!20}     110.769$^\circ$ & \cellcolor{gray!20}      68.571$^\circ$ & \cellcolor{gray!20}      49.655$^\circ$ & \cellcolor{gray!20}      38.919$^\circ$ & \cellcolor{gray!20}      32.000$^\circ$ & \cellcolor{gray!20}      27.170$^\circ$ & \cellcolor{gray!20}      23.607$^\circ$ \\
 & & {\color{gray!70}$G_{6}^1 / G_{7}^N $} &  {\color{gray!70}   8 /   13} &  {\color{gray!70}   8 /   21} &  {\color{gray!70}   8 /   29} &  {\color{gray!70}   8 /   37} &  {\color{gray!70}   8 /   45} &  {\color{gray!70}   8 /   53} &  {\color{gray!70}   8 /   61} \\\cline{2-10}
 & \multirow{2}{*}{8} & \cellcolor{gray!20} $\psi_{8}^N$ & \cellcolor{gray!20}     111.429$^\circ$ & \cellcolor{gray!20}      68.824$^\circ$ & \cellcolor{gray!20}      49.787$^\circ$ & \cellcolor{gray!20}      39.000$^\circ$ & \cellcolor{gray!20}      32.055$^\circ$ & \cellcolor{gray!20}      27.209$^\circ$ & \cellcolor{gray!20}      23.636$^\circ$ \\
 & & {\color{gray!70}$G_{7}^1 / G_{8}^N $} &  {\color{gray!70}  13 /   21} &  {\color{gray!70}  13 /   34} &  {\color{gray!70}  13 /   47} &  {\color{gray!70}  13 /   60} &  {\color{gray!70}  13 /   73} &  {\color{gray!70}  13 /   86} &  {\color{gray!70}  13 /   99} \\\cline{2-10}
 & \multirow{2}{*}{9} & \cellcolor{gray!20} $\psi_{9}^N$ & \cellcolor{gray!20}     111.176$^\circ$ & \cellcolor{gray!20}      68.727$^\circ$ & \cellcolor{gray!20}      49.737$^\circ$ & \cellcolor{gray!20}      38.969$^\circ$ & \cellcolor{gray!20}      32.034$^\circ$ & \cellcolor{gray!20}      27.194$^\circ$ & \cellcolor{gray!20}      23.625$^\circ$ \\
 & & {\color{gray!70}$G_{8}^1 / G_{9}^N $} &  {\color{gray!70}  21 /   34} &  {\color{gray!70}  21 /   55} &  {\color{gray!70}  21 /   76} &  {\color{gray!70}  21 /   97} &  {\color{gray!70}  21 /  118} &  {\color{gray!70}  21 /  139} &  {\color{gray!70}  21 /  160} \\\cline{2-10}
 & \multirow{2}{*}{10} & \cellcolor{gray!20} $\psi_{10}^N$ & \cellcolor{gray!20}     111.273$^\circ$ & \cellcolor{gray!20}      68.764$^\circ$ & \cellcolor{gray!20}      49.756$^\circ$ & \cellcolor{gray!20}      38.981$^\circ$ & \cellcolor{gray!20}      32.042$^\circ$ & \cellcolor{gray!20}      27.200$^\circ$ & \cellcolor{gray!20}      23.629$^\circ$ \\
 & & {\color{gray!70}$G_{9}^1 / G_{10}^N $} &  {\color{gray!70}  34 /   55} &  {\color{gray!70}  34 /   89} &  {\color{gray!70}  34 /  123} &  {\color{gray!70}  34 /  157} &  {\color{gray!70}  34 /  191} &  {\color{gray!70}  34 /  225} &  {\color{gray!70}  34 /  259} \\\cline{2-10}
 & \multirow{2}{*}{11} & \cellcolor{gray!20} $\psi_{11}^N$ & \cellcolor{gray!20}     111.236$^\circ$ & \cellcolor{gray!20}      68.750$^\circ$ & \cellcolor{gray!20}      49.749$^\circ$ & \cellcolor{gray!20}      38.976$^\circ$ & \cellcolor{gray!20}      32.039$^\circ$ & \cellcolor{gray!20}      27.198$^\circ$ & \cellcolor{gray!40} \textbf{     23.628}$^\circ$ \\
 & & {\color{gray!70}$G_{10}^1 / G_{11}^N $} &  {\color{gray!70}  55 /   89} &  {\color{gray!70}  55 /  144} &  {\color{gray!70}  55 /  199} &  {\color{gray!70}  55 /  254} &  {\color{gray!70}  55 /  309} &  {\color{gray!70}  55 /  364} &  {\color{gray!70}  55 /  419} \\\cline{2-10}
 & \multirow{2}{*}{12} & \cellcolor{gray!20} $\psi_{12}^N$ & \cellcolor{gray!20}     111.250$^\circ$ & \cellcolor{gray!20}      68.755$^\circ$ & \cellcolor{gray!20}      49.752$^\circ$ & \cellcolor{gray!40} \textbf{     38.978}$^\circ$ & \cellcolor{gray!20}      32.040$^\circ$ & \cellcolor{gray!40} \textbf{     27.199}$^\circ$ & \cellcolor{gray!40} \textbf{     23.628}$^\circ$ \\
 & & {\color{gray!70}$G_{11}^1 / G_{12}^N $} &  {\color{gray!70}  89 /  144} &  {\color{gray!70}  89 /  233} &  {\color{gray!70}  89 /  322} &  {\color{gray!70}  89 /  411} &  {\color{gray!70}  89 /  500} &  {\color{gray!70}  89 /  589} &  {\color{gray!70}  89 /  678} \\\cline{2-10}
 & \multirow{2}{*}{13} & \cellcolor{gray!20} $\psi_{13}^N$ & \cellcolor{gray!20}     111.245$^\circ$ & \cellcolor{gray!40} \textbf{     68.753}$^\circ$ & \cellcolor{gray!40} \textbf{     49.750}$^\circ$ & \cellcolor{gray!40} \textbf{     38.977}$^\circ$ & \cellcolor{gray!40} \textbf{     32.040}$^\circ$ & \cellcolor{gray!40} \textbf{     27.198}$^\circ$ & \cellcolor{gray!40} \textbf{     23.628}$^\circ$ \\
 & & {\color{gray!70}$G_{12}^1 / G_{13}^N $} &  {\color{gray!70} 144 /  233} &  {\color{gray!70} 144 /  377} &  {\color{gray!70} 144 /  521} &  {\color{gray!70} 144 /  665} &  {\color{gray!70} 144 /  809} &  {\color{gray!70} 144 /  953} &  {\color{gray!70} 144 / 1097} \\\cline{2-10}
 & \multirow{2}{*}{14} & \cellcolor{gray!20} $\psi_{14}^N$ & \cellcolor{gray!40} \textbf{    111.247}$^\circ$ & \cellcolor{gray!20}      68.754$^\circ$ & \cellcolor{gray!40} \textbf{     49.751}$^\circ$ & \cellcolor{gray!40} \textbf{     38.978}$^\circ$ & \cellcolor{gray!40} \textbf{     32.040}$^\circ$ & \cellcolor{gray!40} \textbf{     27.198}$^\circ$ & \cellcolor{gray!40} \textbf{     23.628}$^\circ$ \\
 & & {\color{gray!70}$G_{13}^1 / G_{14}^N $} &  {\color{gray!70} 233 /  377} &  {\color{gray!70} 233 /  610} &  {\color{gray!70} 233 /  843} &  {\color{gray!70} 233 / 1076} &  {\color{gray!70} 233 / 1309} &  {\color{gray!70} 233 / 1542} &  {\color{gray!70} 233 / 1775} \\
\end{tabular}
	}
\end{table}

\FloatBarrier
\newpage

\subsection{RAGA Sampling}

\subsubsection*{Combining Equidistant Angular and Golden Ratio Sampling}

By using a rational approximation for the angle $\psi^N$ 
in Equation~\eqref{Eq::raga_approx},
the sampling scheme is fully defined in terms of the two integers
$G_i^N$ and $G_{i-1}^1$. Here, $G_i^N$  defines a fundamental base angle
\begin{align}
	\phi_i^N := \frac{\pi}{G_i^N}
	\label{Eq::equi-dist-half}
\end{align}
as an integer fraction of the half-circle and $G_{i-1}^1$
defines an index increment between consecutive spokes.
Multiples of the base angle generate exactly the same angular
positions that also occur in an equidistant angular sampling scheme
with $S := G_i^N$ spokes.

\begin{figure}[!ht]
	\centering
	\includegraphics[width=0.8\textwidth]{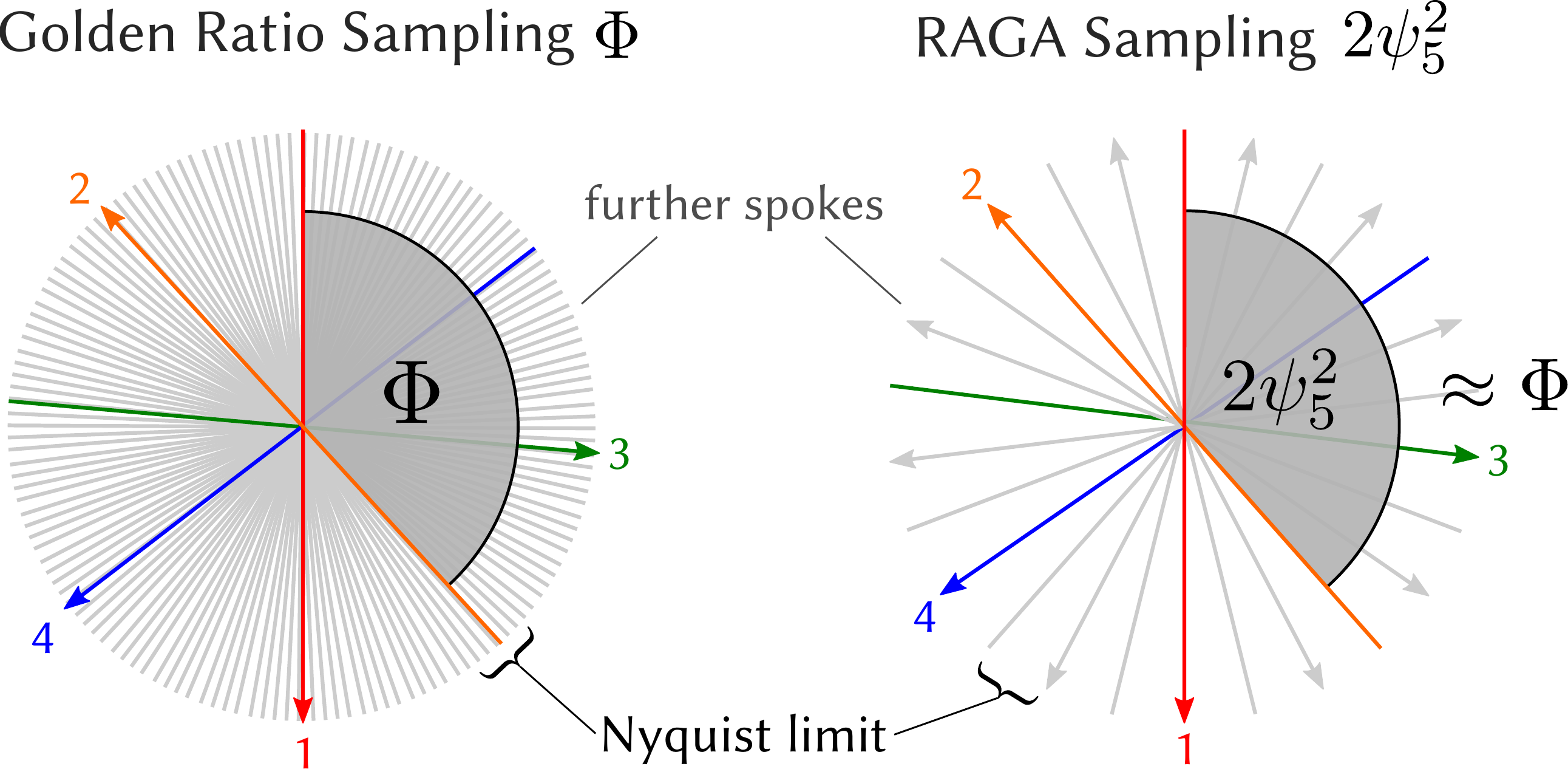}
	\caption{$~$Comparison of a golden ratio (left) and a
		RAGA (right) sampling scheme.
		The golden ratio sampling scheme uses the golden angle $\Phi$.
		The first four spokes of the trajectory are highlighted.
		Because the angle is irrational the spoke angles never repeat and
		new projections are acquired each time.
		The RAGA sampling scheme approximates $\Phi$ with an approximation order of $i=5$.
		It acquires the same set of spokes as an equidistant angular pattern with 13 spokes,
		but reordered so that the angle
		between temporally consecutive spokes approximates an angle of $\Phi$.
		The pattern repeats after all 13 spokes of the equidistant pattern.
		All spokes are acquired exactly once. Note that a low number of 13 spokes
		was used for illustration only, practical RAGA trajectories  would use
		a higher number of spokes corresponding to the Nyquist limit.
		}
	\label{Fig:figure_01}
\end{figure}

By choosing the increment between two temporally consecutive spokes
as $G_{i-1}^1$, the angle between both spokes is
\begin{align}
	\psi^N \approx G_{i-1}^1 \cdot \phi_i^N~.
\end{align}
Thus, in RAGA sampling the temporal scheme is close to a golden ratio
sampling scheme, but all acquired samples are members of an
equidistant angular pattern.
The index of the acquired spoke in the equidistant angular pattern $\text{ind}_t$
is calculated with the temporal sampling index $t$ according to
\begin{align}
	\mathbb{Z} \rightarrow \mathbb{Z}/S\mathbb{Z}, \quad
	t \mapsto \text{ind}_t := (t\cdot G_{i-1}^1) \mod S,
	\label{Eq::raga_increment}
\end{align}
where $\mathbb{Z} / S\mathbb{Z}$ is the additive group of integers modulo $S$.
An illustration is provided in Figure \ref{Fig:figure_01}.A.
This index encodes the information about the projection angle
of each acquired spoke and can be used to
reorder the equidistant angular data following
their temporal acquisition order.

\subsubsection*{Bijectivity of RAGA Sampling}

When acquiring spokes of an equidistant angular pattern with the increment $G_{i-1}^1$ following 
Equation \eqref{Eq::raga_increment} the mapping from $t = 0, \cdots S - 1$ 
to $\mathbb{Z}/ S\mathbb{Z}$ should be bijective. If this were not the case,
some spokes of the equidistant angular pattern would be sampled multiple times
or not at all, which would mean that the property of the golden ratio sampling
to fill each bin in an optimal way is not preserved. On the other hand,
if it is bijective, then the set of $S$ consecutive spokes of the RAGA
sampling covers all positions in the equidistant scheme with $S$ spokes. 
We therefore call such a bin with size $S$ a full frame.  Even with more spokes
acquired, no new spoke positions will be covered.
Thus, $S$ should be chosen large
enough so that the Nyquist criterion is fulfilled everywhere in k-space for a full frame,
which requires $S \geq \frac{\pi}{2} m$ where $m$ is the matrix size.
Acquiring more spoke positions beyond the Nyquist limit would not provide
more information even for golden ratio sampling,
and a larger bin size would simply average information.
This is demonstrated in Figure \ref{Fig:figure_05}.A.

To prove that we always obtain such a full frame after acquiring $S$ spokes, we
have to show that the increment $G_{i-1}^1$ is a generator of the
cyclic additive group $\mathbb{Z}/ S \mathbb{Z}$, i.e. that its multiples
(modulo $S$) generate all elements of $\mathbb{Z}/ S\mathbb{Z}$.
In other words, $G_{i-1}^1$ has order $S = G_i^N$ and this means 
that the pattern repeats exactly after $S$ spokes.
This is the case when the greatest common denominator of
$G_{i-1}^1$ and $G_i^N$ is one, i.e. $\gcd(G_{i-1}^1, G_i^N) = 1$. 
The proof that the RAGA increment $G_{i-1}^1$ has this property
can be found in Section \ref{Sec::biject_proof}.

\subsubsection*{Approximating Golden Ratio Angles}

\begin{figure}[!ht]
	\centering
	\includegraphics[width=0.9\textwidth]{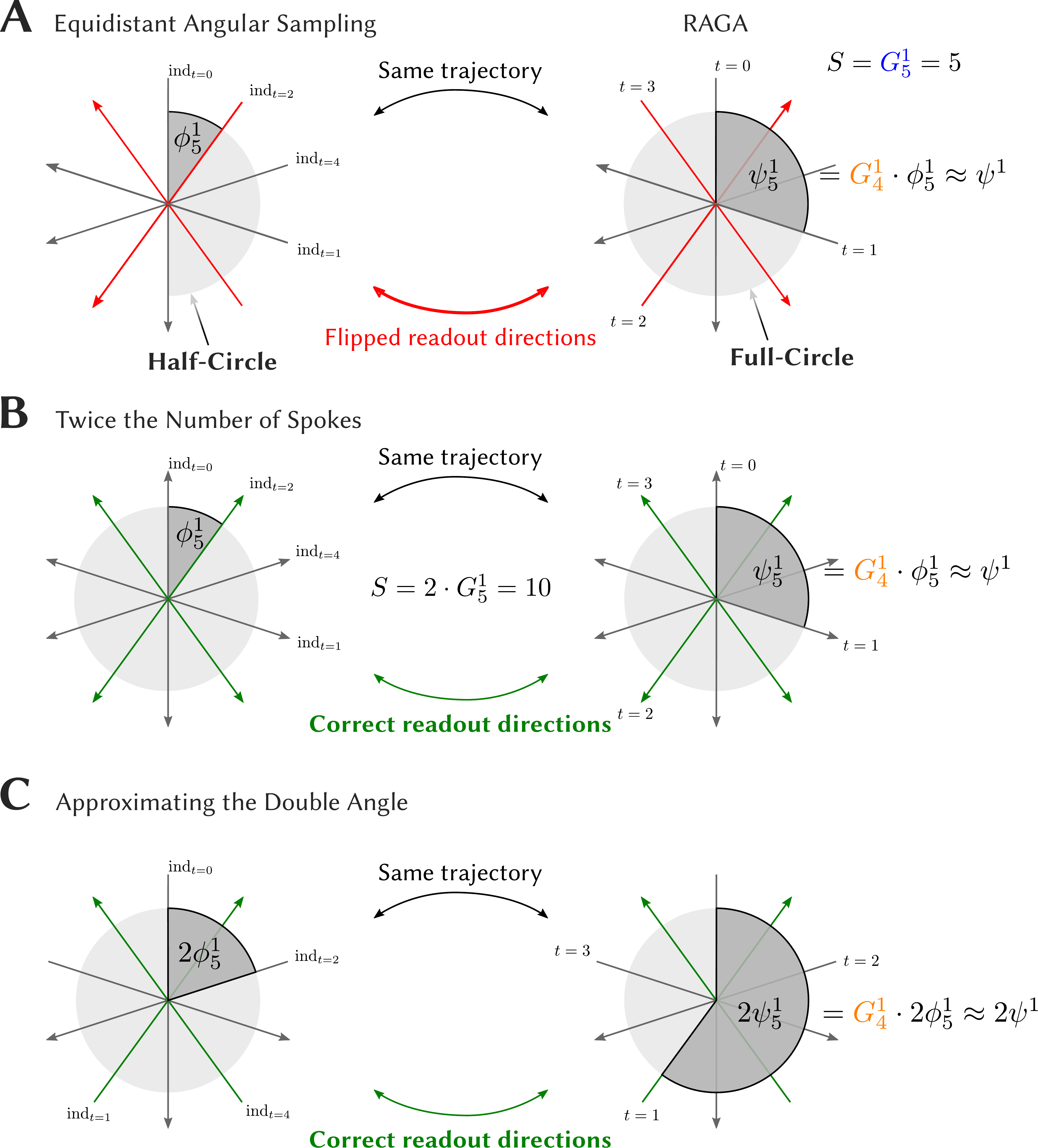}
	\caption{\textbf{A}: An example for an equidistant angular sampling scheme with
		${\color{blue}G_5^1} =$
		5 spokes defined over
		a half-circle (left) and the corresponding RAGA sampling that approximates $\psi^1$ with $\psi_5^1$ (right).
		The sampling order in RAGA corresponds to a golden ratio sampling scheme and
		its temporal evolution is marked with the time index $t$.
		The corresponding indices in the equidistant angular pattern $\text{ind}_t$ are
		calculated with the RAGA increment $G_{i-1}^1=$ 
		${\color[rgb]{1,0.5,0}G_4^1}=$
		$3$ and Equation \eqref{Eq::raga_increment}.
		Extending RAGA to the full circle leads to flipped readout directions
		relative to the equidistant sampling defined over a half-circle.
		This encoding ambiguity can be avoided by either using an extended index space
		or by directly covering a full circle using doubled golden ratio angles.
		\textbf{B}: RAGA sampling using an extended space of indices.
		Golden ration sampling with $\psi^1$ is approximated with RAGA with $\psi_5^1$
		by sampling a full frame with an even number $S = 2 G_i^N$ of spokes and using the
		increment $G_{i-1}^1$.
		\textbf{C}: RAGA sampling approximating the doubled golden ratio angle $2 \psi^1$ using
		an odd number of spokes $S$.
		}
	\label{Fig:figure_02}
\end{figure}

In golden ratio sampling, the sampling schemes are designed to cover a half-circle
in an optimal way, because opposing spokes would not provide
different information.
In practice, the individual spokes are usually distributed around
a full circle which improves robustness by averaging motion and
other inconsistencies \cite{Block_J.KoreanSoc.Magn.Reson.Med._2014}.
Additionally, having opposing spokes can be helpful for
gradient delay correction \cite{Block__2011,Rosenzweig_Magn.Reson.Med._2018a}.
If the corresponding equidistant angular pattern indexed by $0, \cdots, S - 1$
is defined to sample only from one half of the circle, this
leads to flipped readouts when sorting the data according to
Equation \eqref{Eq::raga_increment} as illustrated in Figure \ref{Fig:figure_02}.A.
This problem can  be avoided by extending the space of
indices to $0, \cdots, 2 G_i^N - 1$ to cover a full circle,
reducing modulo $2 G_i^N$. Spokes from the second half of the
circle then correspond to an index equal or larger than $G_i^N$
(Figure \ref{Fig:figure_02}.B).
Note that this does not increase the number of acquired spoke
angles because the indices $n$ and $n + G_i^N$ represent
opposing spokes and this has to be taken into account when
applying the Nyquist criterion. In general, when covering
the full circle using $S$ spokes, there are two different
Nyquist limits for the even and the odd case:
\begin{align}
	S \geq
	\begin{cases}
		\frac{\pi}{2} m & ,~S~\text{is odd}\\
		\pi m & ,~S~\text{is even}~.
	\end{cases}
\end{align}

\subsubsection*{Approximating Doubled Golden Ratio Angles}

Compared to sampling with an odd number of spokes, using twice as
many spokes is often undesirable because this wastes space when
using data files or memory buffers that use zero filling.
Hence, we also consider an alternative strategy that directly covers the
full circle by approximating the doubled golden ratio angle $2 \psi^N$
using a $G_i^N$ that is odd, i.e.
\begin{align}
	2 \psi^N \approx 2 \psi_i^N = G_{i-1}^1\cdot 2 \phi_i^N = G_{i-1}^1 \frac{2 \pi}{G_i^N}~.
\end{align}
Here, the angles again correspond directly to the set of indices $0, \cdots, G_i^N - 1$ (Figure \ref{Fig:figure_02}.C).
The first doubled golden ratio angles are listed in Table \ref{Tab:doubled_angles}.\\
Thus, following Equations \eqref{Eq::raga_approx} and \eqref{Eq::raga_increment}
RAGA sampling can be applied to approximate sampling with the original golden ratio angle $\psi^1$,
the tiny golden angles $\psi^N$, the doubled tiny golden angles $2 \psi^N$,
and the Golden Angle $\Phi = 2\pi - 2 \psi^1$.

\begin{table}[!ht]
	\renewcommand{\arraystretch}{1.2}
	\caption{$~$Table listing the first seven single golden ratio angles and the first 14
	doubled golden ratio angles.
	\label{Tab:doubled_angles}}
	\centerline{\begin{tabular}{c|c|c|c|c|c|c}
% r! ./table2
\multicolumn{7}{l}{Golden Ratio Angles [$^{\circ}$]} \vspace*{0.1cm} \\
$\psi^{1}$ & $\psi^{2}$ & $\psi^{3}$ & $\psi^{4}$ & $\psi^{5}$ & $\psi^{6}$ & $\psi^{7}$ \\\hline 
    111.246 &      68.754 &      49.751 &      38.978 &      32.040 &      27.198 &      23.628 \vspace*{0.5cm}\\ 
 \multicolumn{7}{l}{Doubled Golden Ratio Angles [$^{\circ}$]} \vspace*{0.1cm} \\
$2\psi^{1}$ & $2\psi^{2}$ & $2\psi^{3}$ & $2\psi^{4}$ & $2\psi^{5}$ & $2\psi^{6}$ & $2\psi^{7}$ \\\hline 
    222.492 &     137.508 &      99.502 &      77.955 &      64.079 &      54.397 &      47.256 \vspace*{0.2cm} \\
$2\psi^{8}$ & $2\psi^{9}$ & $2\psi^{10}$ & $2\psi^{11}$ & $2\psi^{12}$ & $2\psi^{13}$ & $2\psi^{14}$ \\\hline 
     41.773 &      37.430 &      33.905 &      30.986 &      28.531 &      26.436 &      24.627 \\
	\end{tabular}}
\end{table}

\FloatBarrier

\subsubsection*{Numerical Stability of RAGA Sampling}
In RAGA sampling, the index of the projection angle
in the equidistant angular pattern follows from Equation \eqref{Eq::raga_increment} 
with the integers $t$, $G^1_{i-1}$, and $S$.
Due to exact integer arithmetic the index $\text{ind}_t$ is bitwise reproducible.
Inaccuracies due to floating point arithmetic are therefore only introduced by the
multiplication of the index $\text{ind}_t$ with the base angle $\phi_i^N$ from Equation \eqref{Eq::equi-dist-half}.
This error does not propagate beyond a single repetition of the equidistant pattern
rendering RAGA sampling numerically robust.
In contrast, golden ratio sampling is sensitive to numerical inaccuracies which
accumulate over time.

\subsubsection*{Data Storage and Processing Characteristics}
\begin{figure}[!ht]
	\centering
	\includegraphics[width=.7\textwidth]{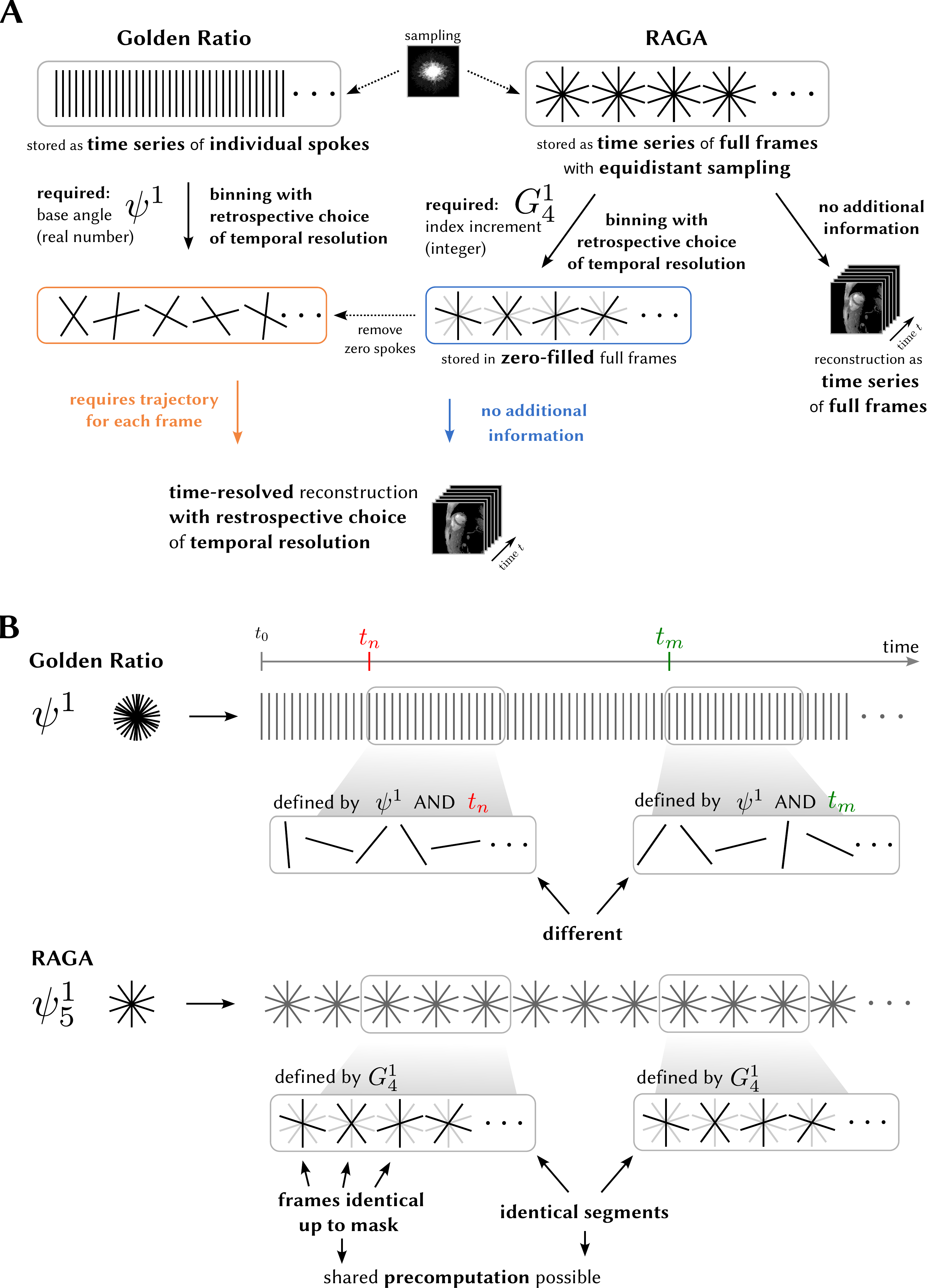}
	\caption{Data storage and processing for RAGA and golden ratio based sampling.
		Note that the small approximation order of $i=5$ is chosen only for illustration purposes.
		\textbf{A}: Golden ratio based data is stored in a time series of individual spokes and knowledge about $\psi^1$ is required for a reconstruction. RAGA sampling can be stored in an equidistant angular order which yields a natural decomposition of the time series into full frames. This allows reconstruction of each full frame even without any additional information. With only knowledge about the index increment $G^1_4$ corresponding to the underlying base angle (here: $\psi^1_5$), the temporal order of all spokes in the RAGA dataset can be recovered and frames with arbitrary temporal footprint can be constructed by binning as in golden ratio sampling.
		\textbf{B}: In RAGA sampling, segments consisting of full frames can be extracted and processed without the need to keep track of their position in the original data set. Furthermore, all full frames have identical spokes which allows sharing of precomputed data for image reconstruction. When extracting data from a non-repeating trajectory, the position $t_n$ needs be known for each fragment to be able to recompute the original trajectory and no precomputed data can be shared.
	}
	\label{Fig:figure_02_5}
\end{figure}

In RAGA sampling all acquired samples are members of an
equidistant angular pattern.
This leads to the flexibility to use either of both
trajectories for image reconstruction
(Figure \ref{Fig:figure_02_5}).
With RAGA ordering,
the data can be flexibly rebinned to retrospectively
select the temporal resolution exactly as in golden ratio schemes.
Reconstructions of golden ratio datasets require that the 
angle between consecutive spokes $\psi^N$ is stored together with the
data. In contrast, for RAGA sampling the angle $\psi_i^N$ can be
recovered from the total number of spokes in a full frame $S$
and the RAGA increment $G_{i-1}^N$.
Both are quantities which are typically stored with the data
similar to the Cartesian case. A full frame of data
can be compactly stored and reconstructed with a simple
equidistant angular scheme. Here, not even the acquisition
index of the spokes is required, and a long time series
naturally decomposes into repeating full frames, which
simplifies extraction and reconstruction of time periods. 
Given the index of the second spoke in a full frame $G_{i-1}^N$,
the angle between temporally
consecutive spokes and the full RAGA scheme can be recovered. 
After rebinning, the selected spokes for each frame can be 
stored in a zero-filled k-space corresponding to a full frame
as illustrated in Figure \ref{Fig:figure_02_5}.

\section{Methods}

\subsection{Implementation}

To investigate the properties of the proposed sampling
numerically and in MRI experiments,
the RAGA sampling scheme was implemented in the 
Berkeley Advanced Reconstruction Toolbox (BART) \cite{Uecker__2013}.
The RAGA sampling was also implemented in a radial sequence
on two 3T systems using IDEA (Siemens Healthcare, Erlangen, Germany).

\subsection{Slidelobe-to-Peak Ratio}

The properties of the PSF
of the RAGA sampling schemes were evaluated by comparing the
slidelobe-to-peak ratio (SPR) to golden ratio trajectories
and to an equidistant angular trajectory.
Trajectories consisting of 754 (for $\psi_{13}^1$), 838 (for $\psi_{10}^7$),
and 419 (for $2\psi_{10}^7$) spokes were studied.
With a base resolution of 200, the Nyquist criterion requires
at least 314 projections.

In a sliding window approach 5 to 60 consecutive spokes $s$ were extracted,
combined into one frame, and the PSF for this frame was calculated by
applying the adjoint non-uniform fast Fourier transformation $\mathcal{\hat{F}}^H$
applied to a vector of ones.
From the PSF, the SPR was calculated as maximum ratio of
off-center $\text{{PSF}}_{\text{off-center}}$ and
center peak value $\text{PSF}_{\text{center}}$, i.e.
\begin{align}
	\text{SPR} = \text{max}\left(\frac{\text{{PSF}}_{\text{off-center}}}{\text{PSF}_{\text{center}}}\right)~.
\end{align}
The temporal evolution of the SPR was calculated for each window.
The time series of SPR values for each window size was
further processed by computing the maximum.
The same analysis was performed to compare the single and double golden ratio angles:
$\psi^1$, $\psi^7$, $2\psi^2$ and $2\psi^{14}$.

\subsection{Precomputed PSF and GROG Gridding}

The computational advantages when using a precomputed
PSF for image reconstruction were evaluated
in a simple numerical phantom experiment.
Simulated datasets were created in frequency domain
for the Shepp Logan phantom using a single coil,
base resolution of 200, 377 radial spokes, and 100 time steps
using a golden ratio angle $\Phi$ and RAGA sampling $2\psi_{12}^2$.
The spokes were binned to 29 spokes per frame
and reconstructed with an inverse nuFFT on a CPU (Intel Xeon Gold 6136 CPU @ 3.00 GHz)
and GPU (Nvidia Tesla V100) using BART with and without Toeplitz embedding.
For use with Toeplitz embedding when calculating the inverse 
nuFFT, the irrational golden ratio sampling required the
calculation of a new PSF in all frames, while
the selected RAGA sampling pattern repeats after 13
frames such that only 13 PSFs have to be calculated.
The analysis was performed 10 times and the mean and standard derivations were calculated.
10 additional simulations with the same parameter settings but 8 simulated coils were performed
to compare the calibration and gridding time for
GRAPPA operator gridding (GROG) preprocessing of both sampling schemes.
The calibration was performed on 300 spokes
and all spokes of both trajectories were gridded.

\subsection{Numerical Stability of Golden Ratio Trajectories}
\label{ssec::num_stab_golden_ratio}
The sensitivity of golden ratio based sampling schemes
to numerical inaccuracies was investigated
numerically using six different implementations. The projection angle
was computed using different numerical implementations 
for $\psi^1$ and up to 500 000 repetitions
corresponding to 16:40 min of continuous
acquisition with a TR of 2 ms.
Implementations using floating point arithmetic with
single (\textbf{single}) and double precision (\textbf{double}) were included
and compared to an implementation using quadruple precision.

\FloatBarrier

The computation of the projection angle over many repetitions $n$ was studied for 
a multiplicative formula $(\cdot)$ following
\begin{align}
	\phi_t = t\cdot\psi^1~~~~~\text{with}~t\in\mathbb{N}_0~,
	\label{eq:multi_lin_inc_traj}
\end{align}
for a formula using an additive increment $(+)$ according to
\begin{align}
	\phi_t =
	\begin{cases}
		\phi_{t-1}+\psi^1 & \, t \in\mathbb{N} \\
		0 & \, t = 0
	\end{cases}~,
	\label{eq:add_lin_inc_traj}
\end{align}
and for an additive increment followed by modular reduction after
each update $(+,\text{mod})$ with
\begin{align}
	\phi_t =
	\begin{cases}
		(\phi_{t-1}+\psi^1)\mod 2\pi & \, t \in\mathbb{N} \\
		0 & \, t = 0
	\end{cases}~.
	\label{eq:mod_lin_inc_traj}
\end{align}
RAGA angles based on $\psi^1_{13}$ are computed 
in single and double precision according to
\begin{align*}
	\phi_n = \text{ind}_t \cdot \frac{\pi}{G^N_{13}}~~~~~\text{with}~t\in\mathbb{N}_0~,
\end{align*}
with $\text{ind}_t$ given by Equation \eqref{Eq::raga_increment}.

\subsection{Phantom Experiment}

% Measurement Information
To experimentally confirm that RAGA has similar properties
to golden ratio sampling when used to retrospectively bin
data, we acquired two phantom data sets. To confirm that
full frames which fulfill the Nyquist criterion
achieve optimal resolution, we acquired
steady state images from a static phantom with a FLASH sequence.
To demonstrate that RAGA has the same properties
with respect to a retrospective choice of temporal resolution,
we also acquired transient magnetization using
inversion-recovery (IR) FLASH.

RAGA data was acquired for a NIST (National Institute of Standards and Technology)
phantom (model 106, $T_1$ sphere) \cite{Stupic_Magn.Reson.Med._2021}
on a Siemens Vida 3T system (Siemens Healthcare, Erlangen, Germany)
using a 20 channel head-coil. 
A 2D FLASH sequence (TR/TE = 3.2/2.04 ms, flip angle: 8$^{\circ}$, 
bandwidth-time-product (BWTP): 1.6,
RF pulse duration:  0.4 ms, base resolution: 256 samples
with two-fold oversampling in readout direction,
FOV: 200$\times$200 mm$^2$)
was used with RAGA sampling schemes that approximate $2\psi_i^1$
with orders $i\in \{7,9,10,12,13,15,16\}$.
Additionally, data was acquired with
2D IR FLASH sequence (TR/TE = 2.9/1.77 ms,
flip angle: 8$^{\circ}$, BWTP: 1.6, RF pulse duration: 0.4 ms, 
base resolution: 200 samples with two-fold oversampling in readout direction,
FOV: 200$\times$200 mm$^2$)
with non-selective inversion and a RAGA sampling scheme approximating
$2\psi_{13}^1$ which corresponds to a number of spokes that
fulfills the Nyquist criterion.

% Analysis
In a preprocessing step the gradient delays were determined
with RING and 
a corrected trajectory is used for reconstruction
\cite{Rosenzweig_Magn.Reson.Med._2018a}.
Coil profiles were calculated with ESPIRiT \cite{Uecker_Magn.Reson.Med._2014}.
The reconstruction was performed with the equidistant angular trajectory
using an iterative SENSE reconstruction using the methods of conjugate
gradients without regularization and a maximum number of 30 iteration steps.

The IR data was compressed from 18 coils to
12 virtual coils using a singular value decomposition (SVD) \cite{Huang_Magn.Reson.Imaging_2008}.
The data was retrospectively binned to 21, 55, and 233 spokes per frame.
All IR datasets were reconstructed backwards in time with
real-time (RT)-NLINV \cite{Uecker_NMRBiomed._2010}.
The first reconstructed frames are then from the steady-state
of the FLASH readouts, which is helpful when using temporal
regularization.
A starting regularization factor $\alpha_0$ of 1,
a reduction factor of $1/2$ for decreasing $\alpha_n$ in each Gauss-Newton step,
a temporal damping factor of 0.9, and 10 iterations were used
for the RT-NLINV reconstruction.

\subsection{In Vivo Experiment}

% Measurement Information
To investigate temporal resolution in a more complex scenario,
short-axis views of a heart were acquired for a healthy volunteer 
in expiration after obtaining written informed consent
and with approval of the local ethics board using a
real-time radial 2D FLASH (TR/TE = 2.9/1.77 ms, flip angle: 8$^{\circ}$, BWTP: 1.6,
RF pulse duration: 0.4 ms, FOV: 320$\times$320 mm$^2$,
base resolution: 200 with two-fold oversampling in readout direction)
with 18 channels of a combined thorax and spine coil
on a Siemens Skyra 3T system.
Two datasets were acquired, one with a golden ratio angle of $\psi^1$ and
one with the corresponding RAGA sampling using $\psi_{13}^1$
with 754 spokes per pattern and 5 repetitions.

% Analysis
In a preprocessing step the data was compressed from 18 coils to
8 virtual coils using an SVD \cite{Huang_Magn.Reson.Imaging_2008}.
Gradient delays were determined and the corrected trajectory is
used for reconstruction.
Image reconstruction was performed with RT-NLINV after
binning the golden ratio and RAGA sampled datasets
retrospectively to 20, 25, 30, and 40 spokes per frame.
A starting regularization factor $\alpha_0$ of 1,
a reduction factor of $1/2$, and
a temporal damping of 0.9 were manually selected for the 8 iterations
of the RT-NLINV reconstruction.
In a post-processing step a temporal median filter was
applied \cite{Uecker_NMRBiomed._2010} and the FOV was
cropped to its central 150$\times$150.

\FloatBarrier
\section{Results}
\subsection{Sidelobe-to-Peak Ratio}

\begin{figure}[!ht]
	\centering
	\includegraphics[width=0.5\textwidth]{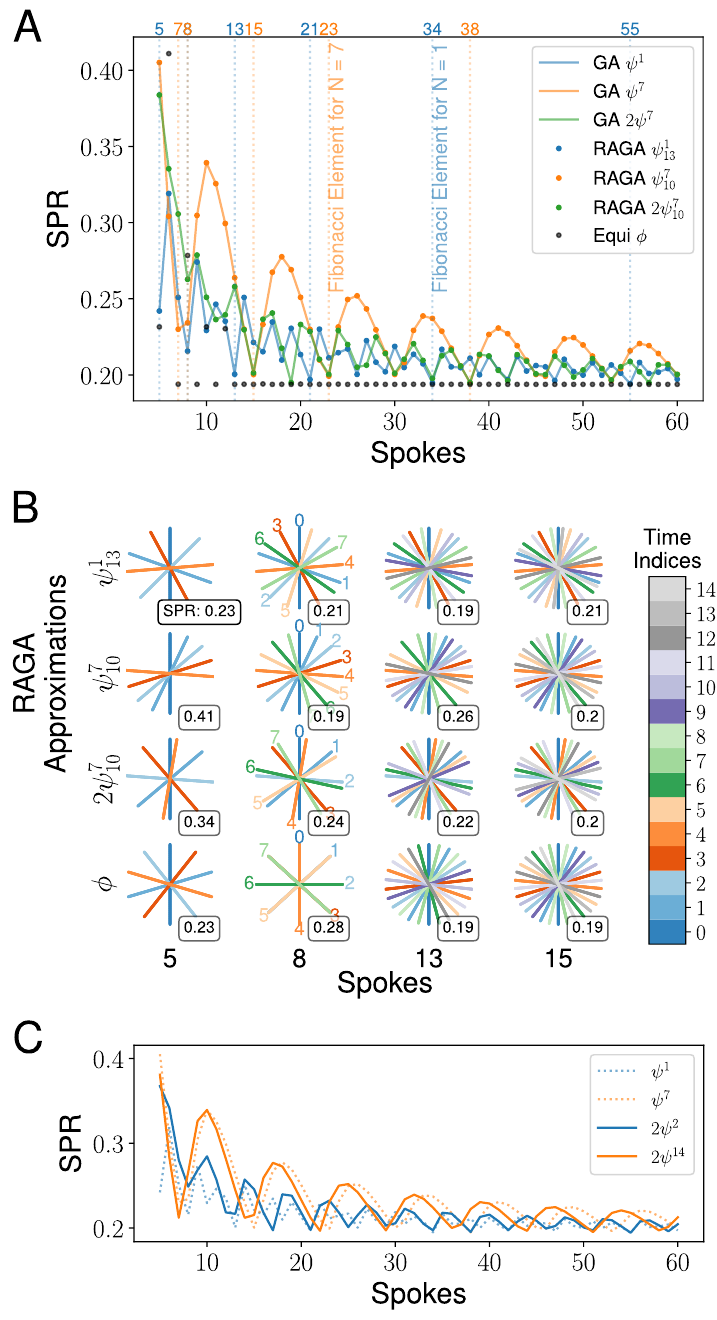}
	\caption{\textbf{A}: Sidelobe-to-peak ratio (SPR) for sampling
	schemes based on the golden ratio angles $\psi^1$, $\psi^7$,
	and $2\psi^7$, their rational approximations with the RAGA sampling
	$\psi_{13}^1$, $\psi_{10}^7$, and $2\psi_{10}^7$, and
	an equidistant scheme using a base angle of $\phi$.
	The SPR is shown for various window sizes.
	Golden ratio angles are plotted with solid lines for better
	differentiation from their rational approximations that are shown with colored dots.
	Elements of the generalized Fibonacci series are marked with dotted vertical lines.
	\textbf{B}: Various sampling schemes for the rational approximations
	$\psi_{13}^1$, $\psi_{10}^7$, and $2\psi_{10}^7$ as well as 
	equidistant sampling with  $\phi$.
	The time index is color-coded and the calculated SPR of
	each sampling scheme is shown.
	\textbf{C}: Calculated SPR for sampling
	schemes based on the single golden ratio angles $\psi^1$ and $\psi^7$ and
	similar doubled golden ratio angles $2\psi^2$ and $2\psi^{14}$.
	The SPR is shown for various window sizes.
	Single golden ratio angles are plotted with dotted lines for better
	differentiation from their doubled alternatives that are shown with solid lines.}
	\label{Fig:figure_03}
\end{figure}

% Description of Figure
Figure \ref{Fig:figure_03}.A shows the SPR values for different window sizes
and sampling schemes for the  angles $\psi^1$, $\psi^7$,
and $2\psi^7$ and their RAGA approximations $\psi_{13}^1$, $\psi_{10}^7$,
and $2\psi_{10}^7$.
For all trajectories the SPR decreases with the number of spokes.
The SPR for the equidistant angular sampling $\phi$
is the lowest for all trajectories except
for small even numbers of spokes, which correspond to less projection angles.
The trajectories based on $\psi^7$ and $2\psi^7$
share the same minima where the number of spokes corresponds to
the elements of the Fibonacci series $G_i^7$, while the SPR values for the trajectory 
based on $\psi^1$ are the smallest
when the number of spokes corresponds to elements of the Fibonacci series $G_i^1$.
The SPR values for $\psi^7$ show periodically higher values compared to the
angles $\psi^1$ and $2\psi^7$ as expected from their
periodically changing homogeneity in k-space coverage.
The single and double golden ratio angles $\psi^1$ and $2\psi^7$
have very similar SPR behavior.
All RAGA approximations very closely match the results of the
corresponding single and doubled golden ratio sampling schemes.

Figure \ref{Fig:figure_03}.B shows the sampling scheme
using $\psi_{13}^1$, $\psi_{10}^7$,
and $2\psi_{10}^7$ for 5, 8, 13, and 15 consecutive spokes.
As expected, the SPR values are large for most cases where 
the k-space coverage is not homogeneous.
For most window sizes the equidistant angular
pattern has the lowest SPR.
This does not hold for small even numbers of spokes
which cover fewer projection angles.

Figure \ref{Fig:figure_03}.C shows the SPR values for different window sizes
and sampling schemes for the single golden ratio angles $\psi^1$, $\psi^7$,
and the similar double golden ratio angles $2\psi^2$ and $2\psi^{14}$.
The same decrease in SPR for larger window sizes and the periodic SPR behavior
as in Figure \ref{Fig:figure_03}.A can be observed.
The single and double golden ratio angles have very similar SPR behavior
which indicates that doubled golden ratio angles could be used
to replace similar single golden ratio angles.

\subsection{Precomputed PSF and GROG Gridding}

\begin{figure}[!ht]
	\centering
	\includegraphics[width=\textwidth]{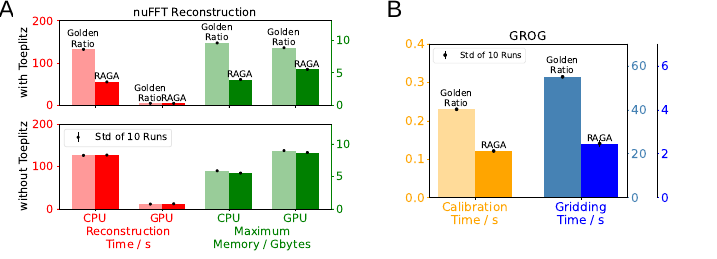}
	\caption{\textbf{A}: Reconstruction times and memory usage of an inverse nuFFT
	averaged over 10 runs with and without
	Toeplitz embedding applied to data of
	the Shepp Logan phantom simulated for trajectories with golden ratio and RAGA sampling scheme.
	\textbf{B}: Calibration and gridding times for data preprocessing using GROG of simulated multi-coil data averaged over 10 runs for golden ratio and RAGA trajectories.}
	\label{Fig:figure_04}
\end{figure}

Figure \ref{Fig:figure_04} shows the mean reconstruction times and maximum of the required memory
of ten runs of an nuFFT reconstruction
with and without Toeplitz embedding, GROG calibration, and GROG gridding of the Shepp Logan phantom
for golden ratio and RAGA sampling.
The computational costs on both CPU and GPU are markedly higher for
the reconstruction of the golden ratio sampled dataset.
The GROG calibration is similarly fast for both techniques,
while the gridding of the golden ratio pattern
is computationally far more expensive compared to the RAGA pattern.

\subsection{Numerical Stability of Golden Ratio Trajectories}

Table \ref{Tab:num_diff} shows the differences in the projection angle resulting from
differing implementations of a golden ratio based sampling scheme
given by Equations \ref{eq:multi_lin_inc_traj}-\ref{eq:mod_lin_inc_traj}.
With the exception of the first repetition, all calculated projection angles
differ from the reference due to the sensitivity of the
golden ratio sampling to numerical inaccuracies.
The differences resulting from floating point arithmetic with single precision
are generally much higher compared to floating point operations with double precision.
The multiplicative formula for the projection angle 
shows the lowest difference to the reference, while the additive formulas
can lead to larger errors already for acquisitions times in the order of seconds
when using only single precision. Avoiding the addition of numbers of different
magnitude with the modular reduction after each update reduced the error markedly.
For RAGA sampling, errors are
on the level of the machine precision even when using single precision and
do not increase over time.

\begin{table}[!ht]
	\renewcommand{\arraystretch}{1.2}
	\caption{$~$
		The projection angles for selected repetitions
		of a continuous golden ratio based sampling scheme with base angle $\psi^1$
		and for the RAGA approximation $\psi^1_{13}$. The
		acquisition times are computed assuming \mbox{TR = 2 ms}.
		Different implementations of the calculation of the projection angle are compared.
		\textbf{S}\textbf{ingle} and \textbf{double} refer to the floating point precision as
		single or double, respectively, and $(\cdot)$, $(+)$, and $(+,\text{mod})$ refer
		to the calculation of the projection angle in time 
		according to Equations \ref{eq:multi_lin_inc_traj}-\ref{eq:mod_lin_inc_traj}.
		Angles and errors were converted to degrees for representation.
		Machine epsilons scaled in the same way are $6.8 \times 10^{-6}$ and $1.3 \times 10^{-14}$ 
		for single and double precision, respectively.
		\label{Tab:num_diff} }
	\centerline{
		\begin{adjustbox}{width=\columnwidth,center}
\begin{tabular}{cl||c|c|c|c|c}
\multicolumn{7}{c}{\hspace{2cm}\textbf{Repetitions}}\\
& & 10000 & 25000 & 100000 & 250000 & 500000\\\cmidrule{2-7}\morecmidrules\cmidrule{2-7}
& {\cellcolor{gray!40}\ \textbf{GA Ref} [$^{\circ}$]} & {\cellcolor{gray!40}61.18} & {\cellcolor{gray!40}152.95} & {\cellcolor{gray!40}251.80} & {\cellcolor{gray!40}89.49} & {\cellcolor{gray!40}178.99}\\\cmidrule{2-7}
& {\cellcolor{gray!40}\  \ \textbf{RAGA Ref} [$^{\circ}$]} & {\cellcolor{gray!40}46.35} & {\cellcolor{gray!40}115.88} & {\cellcolor{gray!40}103.52} & {\cellcolor{gray!40}78.80} & {\cellcolor{gray!40}157.60}\\\cmidrule{2-7}\morecmidrules\cmidrule{2-7}
\multirow{3}{*}{\rotatebox[origin=c]{90}{\textbf{Golden Ratio}}} & {\cellcolor{gray!40}\textbf{single}$_{(\cdot)}$ [$^{\circ}$]}&{\color{gray!70}-2.3e-02}&{\color{gray!70}-1.4e-03}&{\color{gray!70}-5.7e-03}&{\color{gray!70}-4.6e-01}&{\color{gray!70}-9.2e-01}\\\cmidrule{2-7}
& {\cellcolor{gray!40}\textbf{single}$_{(+)}$ [$^{\circ}$]}&{\color{gray!70}-2.7e+02}&{\color{gray!70}-9.5e+01}&{\color{gray!70}1.7e+02}&{\color{gray!70}-1.4e+02}&{\color{gray!70}7.0e+01}\\\cmidrule{2-7}
& {\cellcolor{gray!40}\textbf{single}$_{(+,\text{mod})}$ [$^{\circ}$]}&{\color{gray!70}-1.1e-01}&{\color{gray!70}-2.7e-01}&{\color{gray!70}-1.1e+00}&{\color{gray!70}-2.7e+00}&{\color{gray!70}-5.4e+00}\\\cmidrule{2-7}\morecmidrules\cmidrule{2-7}
& {\cellcolor{gray!40}\textbf{single}$_{\text{RAGA}}$ [$^{\circ}$]}&{\cellcolor{gray!10}\color{gray!70}-3.4e-06}&{\cellcolor{gray!10}\color{gray!70}-9.7e-11}&{\cellcolor{gray!10}\color{gray!70}-3.6e-06}&{\cellcolor{gray!10}\color{gray!70}-4.1e-06}&{\cellcolor{gray!10}\color{gray!70}-8.2e-06}\\\cmidrule{2-7}\morecmidrules\cmidrule{2-7}
\multirow{3}{*}{\rotatebox[origin=c]{90}{\textbf{Golden Ratio}}} & {\cellcolor{gray!40}\textbf{double}$_{(\cdot)}$ [$^{\circ}$]}&{\color{gray!70}1.3e-10}&{\color{gray!70}2.2e-10}&{\color{gray!70}8.6e-10}&{\color{gray!70}4.9e-10}&{\color{gray!70}9.8e-10}\\\cmidrule{2-7}
& {\cellcolor{gray!40}\textbf{double}$_{(+)}$ [$^{\circ}$]}&{\color{gray!70}2.5e-07}&{\color{gray!70}9.4e-07}&{\color{gray!70}4.4e-06}&{\color{gray!70}-1.8e-04}&{\color{gray!70}-5.9e-04}\\\cmidrule{2-7}
& {\cellcolor{gray!40}\textbf{double}$_{(+,\text{mod})}$ [$^{\circ}$]}&{\color{gray!70}7.2e-11}&{\color{gray!70}1.8e-10}&{\color{gray!70}7.2e-10}&{\color{gray!70}1.8e-09}&{\color{gray!70}3.6e-09}\\\cmidrule{2-7}
& {\cellcolor{gray!40}\textbf{double}$_{\text{RAGA}}$ [$^{\circ}$]}&{\cellcolor{gray!10}\color{gray!70}-3.1e-15}&{\cellcolor{gray!10}\color{gray!70}-1.4e-15}&{\cellcolor{gray!10}\color{gray!70}-5.7e-15}&{\cellcolor{gray!10}\color{gray!70}-1.5e-15}&{\cellcolor{gray!10}\color{gray!70}-2.9e-15}\\\cmidrule{2-7}\morecmidrules\cmidrule{2-7}
& Acquisition [min:s]&0:20&0:50&3:20&8:20&16:40\\\cmidrule{2-7}
\end{tabular}
\end{adjustbox}
	}
\end{table}

\subsection{Phantom Experiment}

\begin{figure}[!ht]
	\centering
	\centerline{\includegraphics[width=.68\textwidth]{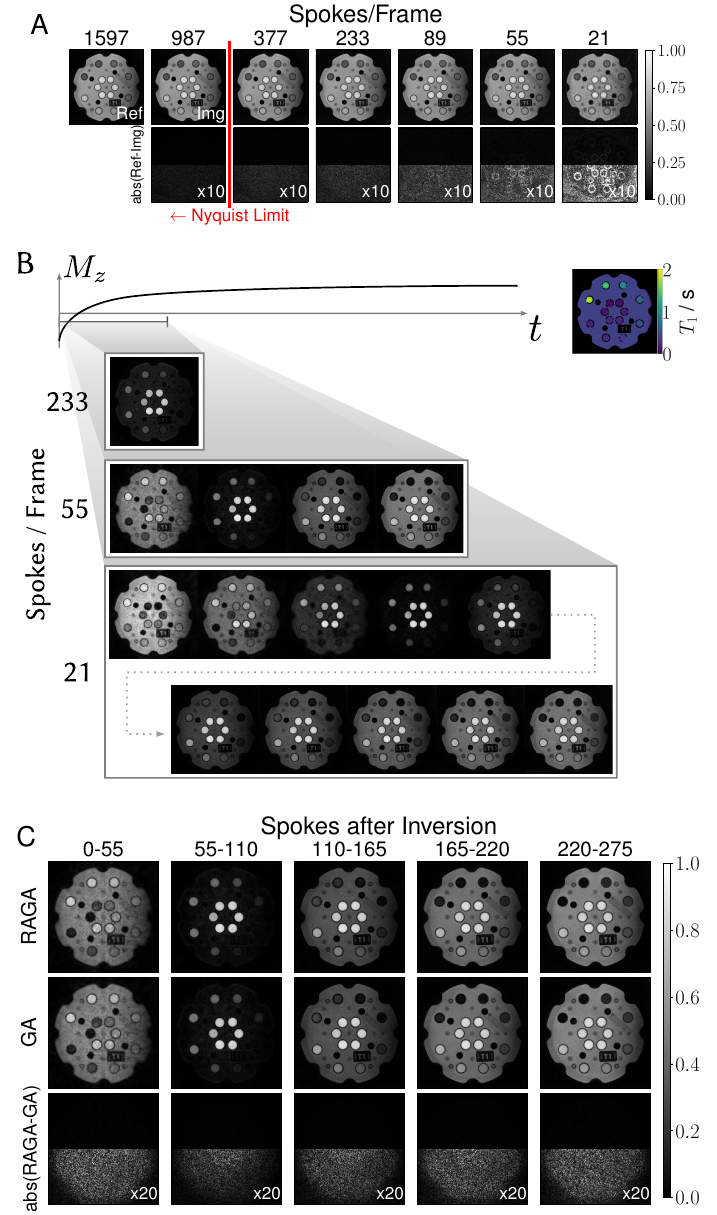}}
	\caption{\textbf{A}: Phantom images acquired with a radial FLASH sequence
	and reconstructed with SENSE
	using a different number of spokes per frame as defined by the approximation
	order of the RAGA sampling for $2\psi_i^1$ with orders
	$i\in \{7,9,10,12,13,15,16\}$.
	Difference maps to the frame with the most spokes are shown, with
	the bottom half scaled by 10.
	\textbf{B}: A series of phantom images acquired with
	an IR FLASH sequence and RAGA sampling with spoke angle $2\psi_{13}^1$.
	The data was rebinned to 21, 55, and 233 spokes per frame
	and reconstructed using RT-NLINV
	demonstrating the increasing temporal resolution.
	A $T_1$ map was calculated from the rebinned
	dataset with the highest temporal resolution, i.e. 21 spokes per frame.
	\textbf{C}: Selected frames from the time series showing the
	inversion recovery acquired with 55 spokes per frame using
	RAGA angle $2\psi_{13}^1$.
	The difference maps show the difference to an acquisition
	with golden ratio angle $2\psi^1$ with the bottom half scaled up by 20.}
	\label{Fig:figure_05}
\end{figure}

Figure \ref{Fig:figure_05}.A shows different reconstructions of a FLASH sequence
acquired with RAGA sampling schemes of varying approximation order.
This results in different numbers of spokes per frame.
The differences of the individual maps with respect to the map with
the most acquired spokes are shown.
The error increases slightly when the number of spokes is reduced,
but does not include any structural details until fewer than 89 spokes per frame
are used.
The Nyquist limit in this experiment corresponds to 402 spokes per frame.
From this point on, increasing the number of projection angles does not
add more information.

The rows in Figure \ref{Fig:figure_05}.B 
show differently binned datasets with varying temporal resolution
covering almost the same time interval during the IR FLASH experiment.
With a reduced number of spokes per frame more states during the 
recovery of the magnetization can be resolved.

Figure \ref{Fig:figure_05}.C shows reconstructions of two IR FLASH experiments
acquired with a golden ratio and RAGA sampling scheme.
Their differences are low in each time intervals during the recovery
and  appear noise-like without structural details. This confirms
that both schemes have the same flexibility in choosing
the temporal resolution retrospectively.

\subsection{In Vivo Experiment}

\begin{figure}[!ht]
	\centering
	\includegraphics[width=\textwidth]{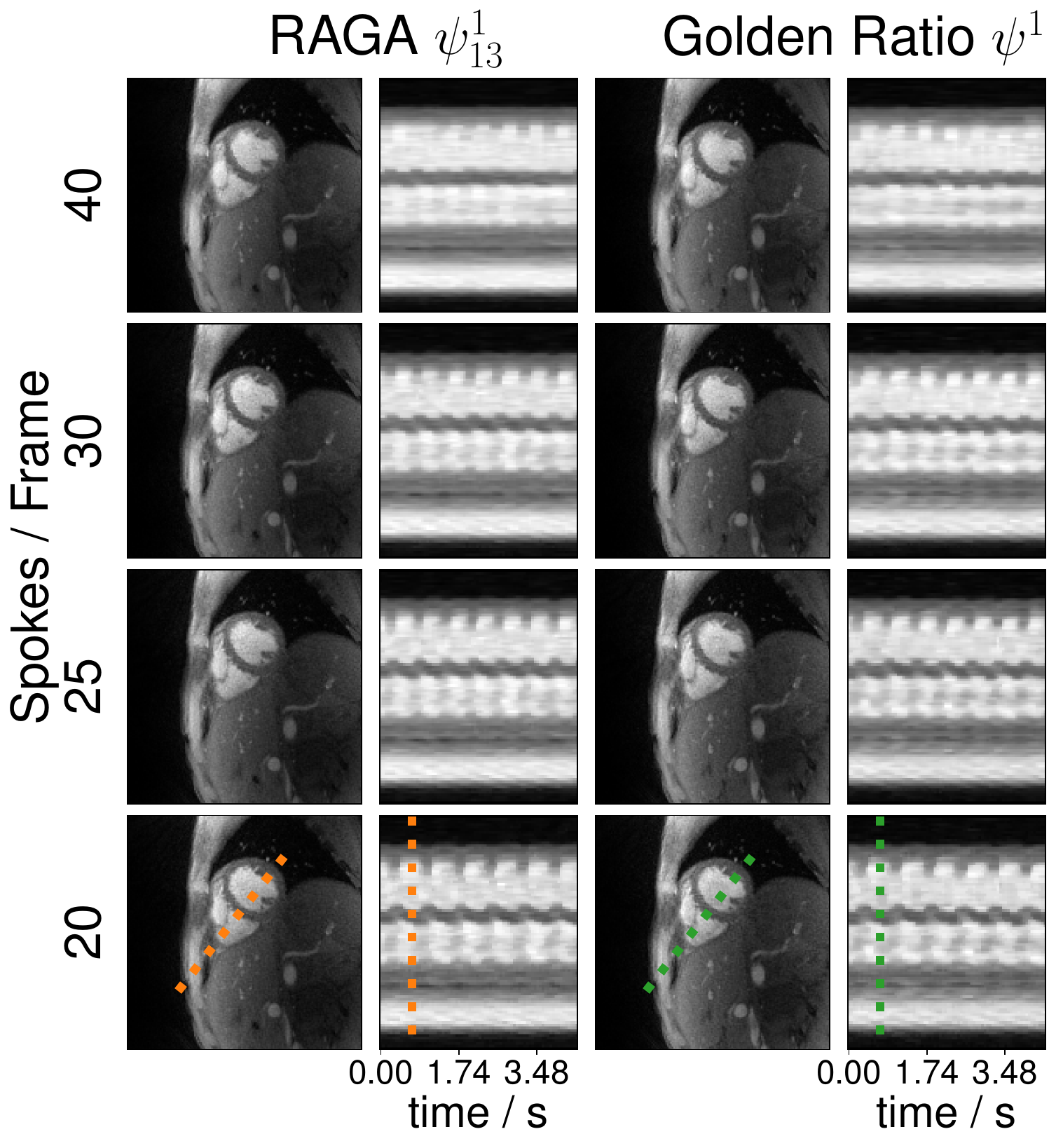}
	\caption{Short-axis views of a human heart acquired with
	real-time radial FLASH using RAGA
	(full frame of 233 spokes)
	and golden ratio sampling
	and reconstructed using RT-NLINV after retrospective binning
	to different temporal resolutions.
	The left two rows show an image from the diastolic phase
	sampled with the rational approximation $\psi_{13}^1$
	and the time evolution of an extracted line profile.
	The right shows the same analysis for the golden ratio angle $\psi^1$.}
	\label{Fig:figure_06}
\end{figure}

Figure \ref{Fig:figure_06} shows the RT-NLINV reconstructions of a
real-time FLASH acquisition of a human heart in short-axis view
for different windows sizes for RAGA and golden ratio sampling.
A small difference in slice localization can be observed between
both acquisition with $\psi_{13}^1$ and $\psi^1$ likely caused
by motion or breathing.
The reconstruction quality of the acquisition with RAGA
approximation $\psi_{13}^1$ and with the irrational angle
$\psi^1$ is visually very similar to that of golden ratio sampling. The
extracted line profiles that show a temporal profile
of the cardiac motion are also very similar for all
temporal resolutions even for time intervals with
fast cardiac motion. In summary, the golden ratio and RAGA sampling
provide essentially the same image quality over multiple heart cycles
and temporal resolutions.

\FloatBarrier
\section{Discussion}

Golden ratio based sampling schemes are well suited for dynamic MRI
because they allow a retrospective choice of temporal resolution.
This work simplifies the concept of golden ratio sampling by
reducing them to a reordering scheme for the simpler equidistant sampling strategy.
This enables the sharing of precomputed information
in reconstruction algorithms, improves numerical robustness,
and simplifies data management.  A crucial insight was that
with the appropriate choice of the spoke increment and total
number of spokes, the RAGA sampling scheme not only approximates
the right angle but also steps through all possible angles
of the underlying equidistant sampling scheme.

Although based on a rational approximation, we show that RAGA
sampling preserves all important properties of golden ratio sampling.
This was demonstrated by calculating and comparing the SPR values of
selected irrational angles with their RAGA approximations.
For spokes following the corresponding generalized Fibonacci series the
SPR values of both schemes are close to the theoretical optimum of
the equidistant angular distribution.
For spokes in between elements of the generalized Fibonacci series
the periodically changing homogeneity in k-space coverage
for small angles leads in both schemes to increased SPR values.
The approximation accuracy increases fast with the approximation order.
For sampling schemes that fulfill the Nyquist criterion for typical
base resolutions, the approximation error is lower than one degree.\\
Because golden ratio sampling is based on a definition
using a half circle, while its practical application then typically
uses angles from a full circle, a one-to-one application of RAGA leads
to flipped readouts. This can be avoided by either using doubled golden
ratio angles as base angles or by using a doubled index space.
Doubled golden ratio angles have very similar SPR behavior to
similar golden ratio angles and provide an alternative that
can be used to distribute the spokes over the full circle
without having to artificially increase the index space.

In contrast to golden ratio sampling schemes, RAGA sampling
schemes repeat after acquiring  a finite number of spokes
as defined by the approximation order.
This avoids the accumulation of numerical errors over multiple
repetitions due to floating point arithmetic
or differing numerical implementations. Single precision arithmetic
has to be avoided for golden ratio sampling due to high numerical
errors. While double precision can be very accurate even for longer scan
times and numerical accuracy is then not a limitation, one has to
avoid numerically poor implementations and agree on a single numerical
implementation for exact reproduction. In contrast, RAGA trajectories are numerically
robust for arbitrarily long scans even when using single precision.
Additionally, the sampling scheme and related quantities such
as its PSF can be precomputed and reused in
Toeplitz-accelerated non-Cartesian
reconstructions,
reducing memory requirements and reconstruction
time.
Methods for accelerating non-Cartesian reconstructions
with preprocessing based on GROG benefit from reduced gridding costs,
because the individual shifting kernels $G_x^{\delta x}$ and $G_y^{\delta y}$
can be reused for samples on repeating trajectories.

Golden ratio sampling acquires new projections of the object
with every spoke for the complete time series. While this seems
to be an advantage, once the Nyquist criterion is exceeded
further projection angles do not add additional information
compared to their repeated variants in RAGA sampling.
This was confirmed in a phantom experiment by using a varying number
of spokes per frame. Their comparison showed that
structural differences in the reconstructions only appeared
below the Nyquist limit. This can be used in RAGA sampling
to automatically determine the required approximation order
by using the Nyquist criterion to determine the maximum number of
useful projection angles, further removing an unnecessary
degree of freedom for the user. The sampling scheme is then fully 
defined by the approximated angle $\phi^N$ and the base resolution,
while the order $i$ is automatically set to the lowest value,
which still fulfills the Nyquist criterion.

Compared to golden ratio sampling, RAGA sampling comes with
the flexibility to use either an equidistant angular or
retrospectively rebinned temporal
trajectory for image reconstruction
as demonstrated for IR-FLASH and a real-time cardiac MRI data set.
The data can be conveniently stored and reconstructed without any
additional information as known from Cartesian data. This can be
expected to become even more useful when sampling schemes are
more complicated with additional encoding dimensions that then
often combine different sampling strategies. 
In addition, navigator lines are often repeated exactly and can simply be
reused in the reconstruction by inserting them into at the
corresponding integer position, while this would require special
handling in conventional golden ratio sampling.
Finally, repeated patterns enable precomputation of interpolation
weights, PSF, subspace coefficients, or other quantities that
can be shared  at the level of full frames or individual spokes. As shown in
this work, this can reduce computational demand substantially.
Thus, RAGA data is almost self-explanatory and simplifies
data management in various ways.

% Extensions
In the future, similar schemes could be developed for
three-dimensional radial trajectories \cite{Chan_Magn.Reson.Med._2009,Piccini_MagnResonMed_2011}.

\section{Conclusion}

In this work, we introduced RAGA sampling schemes as simplified
versions of golden ratio sampling using rational
approximations. We proved mathematically that this
leads to temporal sampling schemes which correspond
to equidistant sampling after reordering, which
allows precomputation in reconstruction algorithms,
improves numerical robustness, and simplifies data management.
At the same time, the approximations preserve all the
practical advantages of golden ration sampling, which
was shown using a numerical analysis and phantom and 
in vivo experiments.

\section{Data Availability}
The tools of this work are implemented in \href{https://github.com/mrirecon/bart}{BART} with commit 79fd4a72. They will be part of the future release following version 0.9.00 of the software.
The scripts to reproduce all the figures and Tables can be found at \href{https://github.com/mrirecon/raga}{Github:mrirecon/raga}.
The required datasets can be downloaded from
Zenodo \href{https://doi.org/10.5281/zenodo.10260250}{(DOI: 10.5281/zenodo.10260250)}.
An interactive tutorial demonstrating the various storage
option accessible with RAGA sampling and illustrated in Figure \ref{Fig:figure_02_5}
can be found on \href{https://github.com/mrirecon/raga-tutorial}{Github:mrirecon/raga-tutorial}.

\section{Acknowledgments}

We appreciate Stefan Wundrak's encouraging comments and input, which
helped to correct a mistake in Table 1.
The authors thank the ISMRM Reproducible Research Study Group for conducting
a code review of Version 1 \href{https://doi.org/10.5281/zenodo.11287833}{(DOI: 10.5281/zenodo.11287833)} of the code
supplied in the Data Availability Statement.
The scope of the code review covered only the code’s ease of download,
quality of documentation, and ability to run,
but did not consider scientific accuracy or code efficiency.

This work was supported by DZHK (German Centre for Cardiovascular Research)
funding code: 81Z0300115,
funded in part by DFG (German Research Foundation) under
Germany's Excellence Strategy - EXC 2067/1- 390729940, and
in part by NIH under grant U24EB029240, R01 NS131948, P41 EB017183,
T32 GM136573, and F30 AG077794.

\section{Conflict of Interest}

The authors declare no potential conflict of interests.

\newpage
\appendix

\section{Bijectivity of RAGA Sampling}
\label{Sec::biject_proof}

\begin{Theorem}\label{Theorem_bijective_mapping}
For all $i, N \in \mathbb{N}$ with $i > 1$
the mapping $t \mapsto \text{ind}_t := t \cdot G_{i-1}^1 \mod G^N_{i}$ 
bijectively allocates all indices $\left[0, G^N_i-1\right]$.
\end{Theorem}
The proof requires two lemmata:

\begin{Lemma} \label{Lemma_FibonacciRecursion}
For $i > 2$, the generalized Fibonacci series can also be expressed as
\begin{align}
	G_i^N = N G_{i-1}^1 + G_{i-2}^1~. \label{Eq:calc_rule_fibonacci}
\end{align}
\end{Lemma}

\noindent\textbf{Proof:}
The assertion is shown via induction over $i$. 
From the definition of the generalized Fibonacci sequence Eq.~\eqref{Eq::gen_fib}
the assertion follows for $i=3$ and $i = 4$: 
\begin{align}
	G_3^N & = G_2^N + G_1^N = N + 1 = N G_2^1 + G_1^1 \quad \textrm{and} \\
	G_4^N & = G_3^N + G_2^N = N + 1 + N = N \cdot 2 + 1 = N G_3^1 + G_2^1~.
\end{align}
Assuming the assertion \eqref{Eq:calc_rule_fibonacci} for $i-1$ and $i$ 
and applying it to both terms on the right side of the definition \eqref{Eq::gen_fib}
of the generalized Fibonacci sequence yields
\begin{align*}
	G_{i+1}^N &= G_i^N + G_{i-1}^N \\
		&= (N G_{i-1}^1 + G_{i-2}^1) + (N G_{i-2}^1 + G_{i-3}^1) \\
		&= N(G_{i-1}^1 + G_{i-2}^1) + (G_{i-2}^1 + G_{i-3}^1) = NG_i^1 + G_{i-1}^1~,
\end{align*}
where the last step uses the definition of the standard Fibonacci sequence. $\qed$

\begin{Lemma}
\label{Lemma_gcd_Fibonacci}
For $i \in \mathbb{N}$, $i > 1$, $\gcd(G_{i-1}^1, G_{i}^N) = 1$.
\end{Lemma}

\noindent\textbf{Proof:}
For $i=2$, we can compute $\gcd(G_1^1, G_2^N) = 1$. For $i > 2$,
we first prove the assertion for $N=1$ by induction over $i$. With
the definition of the Fibonacci sequence and the properties of the
greatest common divisor, we obtain
\begin{align*}
	\gcd(G_{i-1}^1, G_i^1)  = \gcd(G_{i-1}^1, G_{i-1}^1 + G_{i-2}^1) 
				= \gcd(G_{i-1}^1, G_{i-2}^1)~.
\end{align*}
Hence, the assertion also follows for all $i > 2$ by induction.
Using Lemma \ref{Lemma_FibonacciRecursion} we can extend these
results to $N > 1$ by computing
\begin{align*}
	\gcd(G_{i-1}^1, G_i^N) = \gcd(G_{i-1}^1, N G_{i-1}^1+G_{i-2}^1) = \gcd(G_{i-1}^1, G_{i-2}^1) = 1~.
\end{align*}
$\qed$

\noindent\textbf{Proof of Theorem \ref{Theorem_bijective_mapping}}: 
We prove injectivity by contradiction: Assuming there exist
$k, l \in \left[0, G_i^N - 1\right]$ with $k \neq l$ 
that are mapped to the same residue class, i.e.
\begin{align*}
	k \cdot G_{i-1}^1 \equiv l \cdot G_{i-1}^1 \mod G_i^N.
\end{align*}
From Lemma \ref{Lemma_gcd_Fibonacci} we have that $G_{i-1}^1$ and $G_i^N$
are coprime, which then implies $k \equiv l \mod G_i^N$ by the rules of modular arithmetic.
But this contradicts the assumption
and therefore proves injectivity. Surjectivity follows from injectivity by observing
that the cardinality of the domain and the cardinality of the codomain of
the mapping are the same and finite.
$\qed$

\newpage
\bibliographystyle{MRM-AMA}
\bibliography{../radiology.bib}

\FloatBarrier
\newpage
\pagestyle{empty}

\iftoggle{SUPMATERIAL}
{
\includepdf[pages=-]{Supplementary/Suplementary_Material.pdf}
}

\end{document}